\documentclass[lettersize,journal]{IEEEtran}
\usepackage{amsmath,amsfonts,amssymb,amsthm}
\usepackage{algorithmic}
\usepackage{algorithm}
\usepackage{array}
\usepackage[caption=false,font=footnotesize,labelfont=rm,textfont=rm]{subfig}
\usepackage{textcomp}
\usepackage{stfloats}
\usepackage{url}
\usepackage{verbatim}
\usepackage{graphicx}
\usepackage{cite}
\usepackage{color}
\begin{document}

\title{Power-Efficient Full-Duplex Satellite Communications Aided by Movable Antennas}

\author{Lifeng Lin, \IEEEmembership{Graduate Student Member, IEEE},
		Jingze Ding, \IEEEmembership{Graduate Student Member, IEEE},\\
		Zijian Zhou, \IEEEmembership{Member, IEEE},
		and Bingli Jiao, \IEEEmembership{Senior Member, IEEE}
\thanks{This work was supported by National Natural Science Foundation of China under Grant 62171006.  The calculations were supported by the High-Performance Computing Platform of Peking University. \textit{(Corresponding authors: Jingze Ding; Zijian Zhou.)}}
\thanks{Lifeng Lin, Jingze Ding, and Bingli Jiao are with the School of Electronics, Peking University, Beijing 100871, China (e-mail: linlifeng@pku.edu.cn; djz@stu.pku.edu.cn; jiaobl@pku.edu.cn).}
\thanks{Zijian Zhou is with the School of Science and Engineering, The Chinese University of Hong Kong, Shenzhen, Guangdong 518172, China (e-mail: zjzhou@cuhk.edu.cn).}
}
\maketitle

\begin{abstract}
This letter investigates a movable antenna (MA)-aided full-duplex (FD) satellite communication system, where the satellite, equipped with both transmit and receive MAs, serves multiple uplink (UL) and downlink (DL) user terminals (UTs) in FD mode.  Specifically, we formulate a multiobjective optimization problem to minimize the UL and DL transmit powers under imperfect channel state information (CSI).  To jointly optimize the MA positions and transmit powers, we propose a two-loop particle swarm optimization (PSO) algorithm based on a multiobjective optimization framework.  Simulation results show that flexible adjustments of MA positions can effectively reduce the total UL and DL transmit powers, while also alleviating the burden on self-interference (SI) cancellation modules.
\end{abstract}

\begin{IEEEkeywords}
Satellite communications, full-duplex, movable antenna, multiobjective optimization, interference mitigation.
\end{IEEEkeywords}

\section{Introduction}
\IEEEPARstart{F}{or} the next generation of wireless networks, satellite communications are expected to play a crucial role in providing global connectivity, particularly in remote and underserved areas \cite{zhu}. For satellite-ground links, the severe attenuation of high-frequency electromagnetic waves over long propagation distances impairs system performance, rendering low-frequency electromagnetic waves more favorable. Thus, frequency resources are becoming increasingly limited. Moreover, the acceleration of commercial space initiatives has led to a rapid increase in low Earth orbit (LEO) satellites to meet the unprecedented demand from terrestrial devices.  This surge poses a significant challenge for spectrum availability, which is accessed on a ``first-come, first-served'' basis \cite{survey}. Together, these factors exacerbate the demand and competition over the limited frequency resources.

Full-duplex (FD) technology acts as a promising approach to deal with the spectrum scarcity issue in satellite communications \cite{FD_satellite}, as it can theoretically double spectral efficiency by enabling simultaneous transmission and reception over the same frequency band.  Nevertheless, the high transmit power required to overcome signal attenuation over long propagation distances also intensifies co-channel interference (CCI) and self-interference (SI) \cite{SI}.  One possible solution is to reduce the transmit power. However, this may lead to a degradation of data transmission quality \cite{MOOP1}.  For these reasons, movable antennas (MAs) are considered for further reducing the SI while enhancing both uplink (UL) and downlink (DL) signals \cite{MA1}.  Compared to systems with fixed-position antennas (FPAs), the MA system can provide additional degrees of freedom (DoFs) in antenna position optimization \cite{MA2, MA3,MA_add1,MA_add2}.  The advantages of the FD systems aided by MAs, including interference reduction and signal enhancement, have been extensively demonstrated in the existing studies \cite{MA2,Ding1, Ding2}.

In light of the above, this letter proposes an MA-aided FD satellite communication system, where the FD satellite is equipped with both transmit and receive MAs to serve multiple UL and DL user terminals (UTs) simultaneously.  Specifically, to achieve power-efficient resource allocation, we first formulate a multiobjective optimization problem to investigate the trade-off between total UL and DL transmit powers under imperfect channel state information (CSI). Then, a two-loop particle swarm optimization (PSO)-based multiobjective optimization algorithm is proposed to jointly optimize the MA positions and the UL and DL power allocations. Finally, simulation results demonstrate that, thanks to the flexible movements of MAs, the proposed system can save the UL and DL transmit powers and reduce the burden on the SI cancellation modules compared to conventional FPA systems.
\section{System Model and Problem Formulation}\label{2}
\begin{figure}[!t]
	\centering
	\includegraphics[width=0.9\columnwidth]{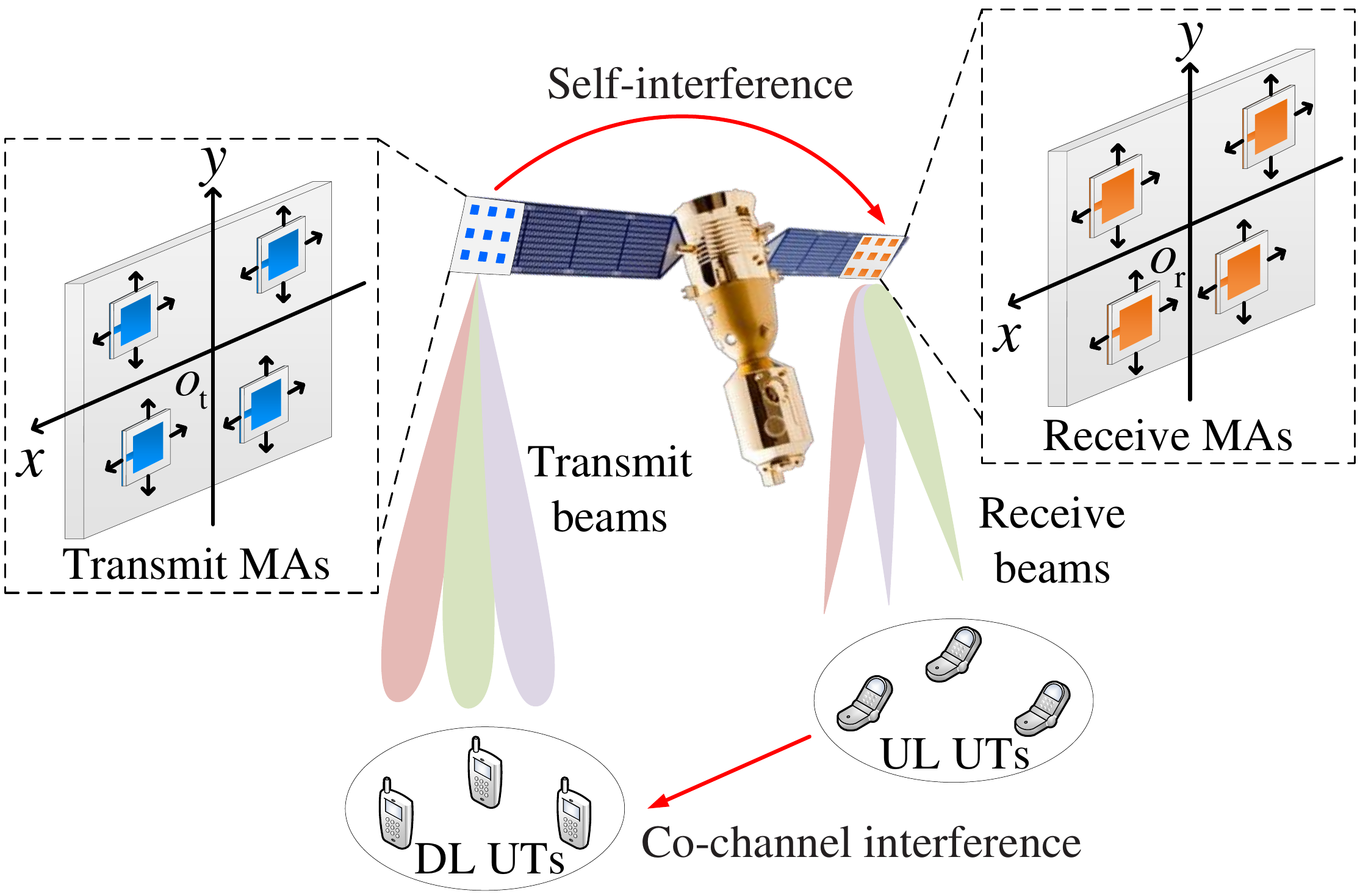}
	\caption{Illustration of the MA-aided FD satellite communication system.}
	\label{system_model}
\end{figure}
\subsection{System Model}
As shown in Fig.\;\ref{system_model}, we consider an MA-aided FD satellite communication system, where an FD satellite, equipped with $M$ transmit MAs and $N$ receive MAs, serves $J$ UL and $K$ DL UTs, each equipped with a single FPA. Each MA is connected to a radio frequency chain via a flexible cable, allowing it to move freely within the designated transmit or receive region, i.e., $\mathcal{C}^\mathrm{t}$ or $\mathcal{C}^\mathrm{r}$. Without loss of generality, $\mathcal{C}^\mathrm{t}$ and $\mathcal{C}^\mathrm{r}$ are assumed to be square regions with sizes $A^\mathrm{t} \times A^\mathrm{t}$ and $A^\mathrm{r} \times A^\mathrm{r}$, respectively. The positions of transmit and receive MAs are described by their Cartesian coordinates, i.e., $\mathbf{t} = {\left[ {\mathbf{t}_1^T, \cdots,\mathbf{t}_m^T,\cdots ,\mathbf{t}_M^T} \right]^T} \in {\mathbb{R}^{2M \times 1}}$ and $\mathbf{r} = {\left[ {\mathbf{r}_1^T, \cdots,\mathbf{r}_n^T,\cdots ,\mathbf{r}_N^T} \right]^T} \in {\mathbb{R}^{2N \times 1}}$, where $\mathbf{t}_m=\left[x_m^\mathrm{t},y_m^\mathrm{t} \right]^T \in \mathcal{C}^\mathrm{t} $ and $\mathbf{r}_n=\left[x_n^\mathrm{r},y_n^\mathrm{r} \right]^T \in \mathcal{C}^\mathrm{r} $, respectively. Due to the mobility of MAs, the SI channel, ${\mathbf{H}_\mathrm{SI}}\left( {\mathbf{t},\mathbf{r}} \right) \in {\mathbb{C}^{N \times M}}$, the UL channel of UT $j$ ($j \in \mathcal{J} \triangleq \left\{ {1, \cdots ,J} \right\}$), ${\mathbf{l}_j}\left( \mathbf{r}\right) \in {\mathbb{C}^{N \times 1}}$, and the DL channel of UT $k$ ($k \in \mathcal{K} \triangleq \left\{ {1, \cdots ,K} \right\}$), ${\mathbf{h}_k}\left(\mathbf{t} \right) \in {\mathbb{C}^{M\times 1}}$, can be actively reconfigured\footnote{Since the satellite's orbit is predetermined, Doppler shifts due to satellite motion can be compensated in advance at the transmitter based on the known Doppler shift distribution in the satellite constellation \cite{survey}. Thus, the proposed channel model primarily focuses on the effects of antenna positioning and omits Doppler shifts.}. 

We consider the field-response channel model \cite{MA1} with the angles of departure (AoDs), angles of arrival (AoAs), and amplitudes of complex coefficients for multiple channel paths do not change regardless of the MA positions, while only the phases vary in the transmit/receive region. The numbers of transmit and receive SI paths are denoted by $L^\mathrm{t}$ and $L^\mathrm{r}$, respectively. The elevation and azimuth AoDs for transmit SI path $l^\mathrm{t}$ ($1 \le l^\mathrm{t} \le L^\mathrm{t}$) and the elevation and azimuth AoAs for receive SI path $l^\mathrm{r}$ ($1 \le l^\mathrm{r} \le L^\mathrm{r}$) are denoted by ${\theta ^\mathrm{t}_{\mathrm{SI},l^\mathrm{t}},\phi ^\mathrm{t}_{\mathrm{SI},l^\mathrm{t}}} \in \left[ {0,\pi } \right]$ and ${\theta ^\mathrm{r}_{\mathrm{SI},l^\mathrm{r}},\phi ^\mathrm{r}_{\mathrm{SI},l^\mathrm{r}}} \in \left[ {0,\pi } \right]$, respectively. Therefore, the signal propagation distance difference of transmit SI path $l^\mathrm{t}$ between position $\mathbf{t}_m$ and the origin of the transmit region, $o_\mathrm{t}$, is $\rho _{\mathrm{SI},l^\mathrm{t}}^\mathrm{t}\left( {{\mathbf{t}_m}} \right) = x_m^\mathrm{t}\sin \theta _{\mathrm{SI},l^\mathrm{t}}^\mathrm{t}\cos \phi _{\mathrm{SI},l^\mathrm{t}}^\mathrm{t} + y_m^\mathrm{t}\cos \theta _{\mathrm{SI},l^\mathrm{t}}^\mathrm{t}$. Similarly, the signal propagation distance difference of receive SI path $l^\mathrm{r}$ between position $\mathbf{r}_n$ and the origin of the receive region, $o_\mathrm{r}$, is $\rho _{\mathrm{SI},l^\mathrm{r}}^\mathrm{r}\left( {{\mathbf{r}_n}} \right) = x_n^\mathrm{r}\sin \theta _{\mathrm{SI},l^\mathrm{r}}^\mathrm{r}\cos \phi _{\mathrm{SI},l^\mathrm{r}}^\mathrm{r} + y_n^\mathrm{r}\cos \theta _{\mathrm{SI},l^\mathrm{r}}^\mathrm{r}$. Let $\lambda$ denote the carrier wavelength. The transmit and receive SI field-response vectors, i.e., ${\mathbf{g}_\mathrm{SI}}\left( {{\mathbf{t}_m}} \right)\in {\mathbb{C}^{{L^\mathrm{t}} \times 1}}$ and ${\mathbf{f}_\mathrm{SI}}\left( {{\mathbf{r}_n}} \right)\in {\mathbb{C}^{{L^\mathrm{r}} \times 1}}$, are then given by
\begin{align}
	{\mathbf{g}_\mathrm{SI}}\left( {{\mathbf{t}_m}} \right) &= {\left[ {{e^{\mathrm{j}\frac{{2\pi }}{\lambda }\rho _{\mathrm{SI},1}^\mathrm{t}\left( {{\mathbf{t}_m}} \right)}}, \cdots ,{e^{\mathrm{j} \frac{{2\pi }}{\lambda }\rho _{\mathrm{SI},L^\mathrm{t}}^\mathrm{t}\left( {{\mathbf{t}_m}} \right)}}} \right]^T}, \\
	{\mathbf{f}_\mathrm{SI}}\left( {{\mathbf{r}_n}} \right) &= {\left[ {{e^{\mathrm{j}\frac{{2\pi }}{\lambda }\rho _{\mathrm{SI},1}^\mathrm{r}\left( {{\mathbf{r}_n}} \right)}}, \cdots ,{e^{\mathrm{j} \frac{{2\pi }}{\lambda }\rho _{\mathrm{SI},L^\mathrm{r}}^\mathrm{r}\left( {{\mathbf{r}_n}} \right)}}} \right]^T} .
\end{align} 
Define the SI channel response from $o_\mathrm{t}$ to $o_\mathrm{r}$ as $\mathbf{\Sigma} \in {\mathbb{C}^{L^\mathrm{r} \times L^\mathrm{t}}}$. The SI channel matrix can be expressed as 
\begin{equation}
	{\mathbf{H}_\mathrm{SI}} \left( { \mathbf{t}}, {\mathbf{r}} \right) ={\mathbf{F}_\mathrm{SI}}{\left( {\mathbf{r}} \right)^H}\mathbf{\Sigma} {\mathbf{G}_\mathrm{SI}}\left( {\mathbf{t}} \right) ,
\end{equation}
where $\mathbf{F}_\mathrm{SI}\left( {\mathbf{r}} \right) = \left[ \mathbf{f}_\mathrm{SI}\left( {\mathbf{r}_1} \right), \cdots, \mathbf{f}_\mathrm{SI}\left( {\mathbf{r}_N} \right) \right] \in {\mathbb{C}^{{L^\mathrm{r}} \times N}}$ and $\mathbf{G}_\mathrm{SI}\left( {\mathbf{t}} \right) = \left[ \mathbf{g}_\mathrm{SI}\left( {\mathbf{t}_1} \right), \cdots, \mathbf{g}_\mathrm{SI}\left( {\mathbf{t}_M} \right) \right] \in {\mathbb{C}^{{L^\mathrm{t}} \times M}}$ are the SI field-response matrices.

For the UL and DL channels in satellite communication systems, the line-of-sight (LoS) path significantly dominates, rendering the non-LoS paths negligible. Therefore, we adopt the LoS channel model for ${\mathbf{l}_j}\left( \mathbf{r}\right)$ and ${\mathbf{h}_k}\left( \mathbf{t}\right)$, i.e.,
\begin{equation}
	{\mathbf{l}_j}\left( \mathbf{r}\right) = \ell_j \tilde{\mathbf{l}}_j\left( \mathbf{r}\right), \quad {\mathbf{h}_k}\left( \mathbf{t}\right) = \hbar_k \tilde{\mathbf{h}}_k \left( \mathbf{t}\right),
\end{equation}
where $\ell_j$ and $\hbar_k$ are the channel coefficients from UL UT $j$ to $o_\mathrm{r}$ and from $o_\mathrm{t}$ to DL UT $k$, and 
\begin{align}
	&\tilde{\mathbf{l}}_j\left( \mathbf{r}\right) = {\left[ {{e^{\mathrm{j}\frac{{2\pi }}{\lambda }\rho _{\mathrm{UL},j}\left( {{\mathbf{r}_1}} \right)}}, \cdots ,{e^{\mathrm{j} \frac{{2\pi }}{\lambda }\rho _{\mathrm{UL},j}\left( {{\mathbf{r}_N}} \right)}}} \right]^T} \in {\mathbb{C}^{N \times 1}}, \\
	& \tilde{\mathbf{h}}_k \left( \mathbf{t}\right)= {\left[ {{e^{\mathrm{j}\frac{{2\pi }}{\lambda }\rho _{\mathrm{DL},k}\left( {{\mathbf{t}_1}} \right)}}, \cdots ,{e^{\mathrm{j} \frac{{2\pi }}{\lambda }\rho _{\mathrm{DL},k}\left( {{\mathbf{t}_M}} \right)}}} \right]^T} \in {\mathbb{C}^{M \times 1}},
\end{align}
are the UL and DL field-response vectors of UL UT $j$ and DL UT $k$, respectively. Here, $\rho _{\mathrm{UL},j}\left( {{\mathbf{r}_n}} \right)= x_n^\mathrm{r}\sin \theta^\mathrm{r}_j\cos \phi^\mathrm{r}_j + y_n^\mathrm{r}\cos \theta^\mathrm{r}_j$ is the signal propagation distance difference of UL UT $j$'s path between position $\mathbf{r}_n$ and $o_\mathrm{r}$, where ${\theta ^\mathrm{r}_j,\phi ^\mathrm{r}_j} \in \left[ {0,\pi } \right]$ are the corresponding elevation and azimuth AoAs, and $\rho _{\mathrm{DL},k}\left( {{\mathbf{t}_m}} \right)= x_m^\mathrm{t}\sin \theta^\mathrm{t}_k\cos \phi^\mathrm{t}_k + y_m^\mathrm{t}\cos \theta^\mathrm{t}_k$ is the signal propagation distance difference of DL UT $k$'s path between position $\mathbf{t}_m$ and $o_\mathrm{t}$, where ${\theta ^\mathrm{t}_k,\phi ^\mathrm{t}_k} \in \left[ {0,\pi } \right]$ are the corresponding elevation and azimuth AoDs.

For the CCI channel, we assume that the FD satellite cannot obtain the perfect CSI\footnote{The CSI of the SI, UL, and DL channels at the FD satellite is assumed to be perfect to provide a performance upper bound for realistic scenarios and robust designs. Despite the challenges in acquiring perfect CSI, existing works \cite{MOOP1,channel_est} have proposed practical methods that provide satisfactory CSI estimation for FD-based or MA-aided systems.}. Based on a deterministic model, the CCI channel from UL UT $j$ to DL UT $k$, denoted by $c_{jk}$, is modeled as
\begin{equation}\label{CCI}
	{c_{jk}} = {{\hat c}_{jk}} + \Delta {c_{jk}} , \quad {\Omega_{jk}} \triangleq \left\{ {{c_{jk}}:\left| {\Delta {c_{jk}}} \right| < {\varepsilon_{jk}}} \right\},
\end{equation}
where ${{\hat c}_{jk}}$ is the estimated CSI, $\Delta {c_{jk}}$ is the estimation error, and ${\Omega_{jk}}$ is a continuous set that contains all possible channel coefficients with a bounded error magnitude of ${\varepsilon_{jk}}$.

For a given time slot, the FD satellite transmits $K$ independent information streams, denoted as $\sum\nolimits_{k \in \mathcal{K}} {\mathbf{w}_k}s_k^\mathrm{DL}$, to $K$ DL UTs, where ${\mathbf{w}_k}\in {\mathbb{C}^{M \times 1}}$ is the beamformer for DL UT $k$ and $s_k^\mathrm{DL}$ is the corresponding DL signal with normalized power. Simultaneously, $J$ UL UTs transmit signals to the FD satellite, denoted as $s_j^\mathrm{UL}$ with normalized power, causing CCI to the DL UTs. Defining the transmit power of UL UT $j$ as $p_j$, the UL and DL receive signals can be expressed as
\begin{align}
&{\mathbf{y}^\mathrm{UL}} = \sum\limits_{j \in \mathcal{J}} {{\mathbf{l}_j}\left( {\mathbf{r}} \right)\sqrt {{p_j}} s_j^\mathrm{UL}}  + {\mathbf{H}_\mathrm{SI}}\left( {\mathbf{t},\mathbf{r}} \right)\sum\limits_{k \in \mathcal{K}} {\mathbf{w}_k}s_k^\mathrm{DL}  + {\mathbf{n}^\mathrm{UL}} , \\
&y_k^\mathrm{DL} = \mathbf{h}_k^H\left( {\mathbf{t}} \right)\sum\limits_{k \in \mathcal{K}} {\mathbf{w}_k}s_k^\mathrm{DL}  + \sum\limits_{j \in \mathcal{J}} {{c_{jk}}\sqrt {{p_j}} s_j^\mathrm{UL}}  + n_k^\mathrm{DL} ,
\end{align}
where ${\mathbf{n}^\mathrm{UL}} \sim \mathcal{CN}\left( {\mathbf{0},\sigma _\mathrm{UL}^2{\mathbf{I}_{{N_\mathrm{r}}}}} \right)$ and ${n_k^{\mathrm{DL}}} \sim \mathcal{CN}\left( {0,\sigma _{\mathrm{DL},k}^2} \right)$ represent the additive white Gaussian noise (AWGN) at the FD satellite and DL UT $k$ with zero means and variances $\sigma _\mathrm{UL}^2$ and $\sigma _{\mathrm{DL},k}^2$, respectively. Denote the zero-forcing receive beamformer for UL UT $j$ as $\mathbf{b}_j=\left(\mathbf{z}_j\left( \mathbf{L}^H \mathbf{L}\right)^{-1}\mathbf{L}^H  \right)^H \in {\mathbb{C}^{N \times 1}}$, where ${\mathbf{z}_j} = [ \underbrace {0, \cdots ,0}_{j - 1},1,\underbrace {0, \cdots ,0}_{J - j}] $ and $\mathbf{L}=\left[{\mathbf{l}_1}\left( {\mathbf{r}} \right), \cdots,{\mathbf{l}_J}\left( {\mathbf{r}} \right) \right] \in {\mathbb{C}^{N \times J}}$, and the SI loss coefficient as $\rho$, which represents the path loss and the SI cancellations in the analog and digital domains\footnote{The movements of the MAs lead to dynamic changes in the SI channel, requiring the SI cancellation modules to adapt efficiently. One potential solution is to integrate machine learning algorithms that can dynamically learn the characteristics of the varying SI channel, thus achieving excellent SI cancellations in the analog and digital domains.}. Define $\mathbf{L}_j\triangleq{\mathbf{l}_j}\left( {\mathbf{r}} \right){\mathbf{l}_j^H}\left( {\mathbf{r}} \right) \in \mathbb{C}^{N \times N}$, $\mathbf{B}_j\triangleq\mathbf{b}_j \mathbf{b}_j^H \in \mathbb{C}^{N \times N}$, $\mathbf{H}_k\triangleq{\mathbf{h}_k}\left( {\mathbf{t}} \right){\mathbf{h}_k^H}\left( {\mathbf{t}} \right) \in \mathbb{C}^{M \times M}$, $\mathbf{W}_k\triangleq\mathbf{w}_k \mathbf{w}_k^H \in {\mathbb{C}^{M \times M}}$, and $S_j\triangleq\rho \mathrm{Tr}\left\lbrace \mathbf{B}_j\mathrm{diag}\left(\sum\nolimits_{k \in \mathcal{K}}{\mathbf{H}_\mathrm{SI}}\left( {\mathbf{t},\mathbf{r}} \right)\mathbf{W}_k{\mathbf{H}_\mathrm{SI}^H}\left( {\mathbf{t},\mathbf{r}} \right) \right)  \right\rbrace $, where $\mathrm{diag}\left(\mathbf{X} \right) $ denotes a diagonal matrix having the main diagonal elements of $\mathbf{X}$ on its main diagonal. Therefore, the achievable rates of UL UT $j$ and DL UT $k$ are obtained as $R^\mathrm{UL}_j=\log_2\left( 1+\gamma^\mathrm{UL}_j\right) $ and $R^\mathrm{DL}_k=\log_2\left( 1+\gamma^\mathrm{DL}_k\right) $, where $\gamma^\mathrm{UL}_j$ and $\gamma^\mathrm{DL}_k$ are the receive signal-to-interference-plus-noise ratios (SINRs) in bits per second per Hertz (bps/Hz) and given by 
\begin{align}
	& \gamma _j^\mathrm{UL} = \frac{{\mathrm{Tr}\left\{ {{\mathbf{L}_j}{\mathbf{B}_j}} \right\}{p_j}}}{{\sum\limits_{i \in \mathcal{J}\backslash \left\{ j \right\}} {\mathrm{Tr}\left\{ {{\mathbf{L}_i}{\mathbf{B}_j}} \right\}{p_i}}  + S_j + \mathrm{Tr}\left\{ {{\mathbf{B}_j}} \right\}\sigma _\mathrm{UL}^2}},\label{gamma_UL} \\
	& \gamma _k^\mathrm{DL} = \frac{{\mathrm{Tr}\left\{ {{\mathbf{W}_k}{\mathbf{H}_k}} \right\}}}{\sum\limits_{i \in \mathcal{K}\backslash \left\{ k \right\}} {\mathrm{Tr}\left\{ {{\mathbf{W}_i}{\mathbf{H}_k}} \right\}}  + \sum\limits_{j \in \mathcal{J}} {{{\left| {{c_{jk}}} \right|}^2}{p_j}}  + \sigma _{\mathrm{DL},k}^2}.\label{gamma_DL}
\end{align} 
\subsection{Problem Formulation}
We first formulate the total UL and DL transmit power minimization problems, and then investigate the joint optimization of the two problems. The total UL transmit power minimization can be formulated as
\begin{align}\label{P1}
	&\mathop {\mathrm{minimize}}\limits_{{p_j},{\mathbf{w}_k}, \mathbf{t},\mathbf{r} } \quad
	\sum\limits_{j \in \mathcal{J}} p_j\\
	&\text{s.t.} \quad \text{C1}: R_j^\mathrm{UL} \ge R_{\mathrm{TH},j}^\mathrm{UL}, \quad \forall j \in \mathcal{J} , \nonumber\\
	&\hspace{2.2em} \text{C2}: \mathop {\min }\limits_{{c_{jk}} \in {\Omega _{jk}}} R_k^\mathrm{DL} \ge R_{\mathrm{TH},k}^\mathrm{DL}, \quad \forall k \in \mathcal{K}, \nonumber\\
	&\hspace{2.2em} \text{C3}: p_j \ge 0, \quad \forall j \in \mathcal{J}, \nonumber\\
	&\hspace{2.2em} \text{C4}: \mathbf{t} \in {\mathcal{C}^\mathrm{t}}, \quad \mathbf{r} \in {\mathcal{C}^\mathrm{r}}, \nonumber\\
	&\hspace{2.2em} \text{C5}: {\left\| {{\mathbf{t}_a} - {\mathbf{t}_{\tilde a}}} \right\|_2} \ge D ,\quad 1 \le a \ne {\tilde a} \le M, \nonumber\\
	&\hspace{2.2em} \text{C6}: {\left\| {{\mathbf{r}_b} - {\mathbf{r}_{\tilde b}}} \right\|_2} \ge D, \quad 1 \le b \ne {\tilde b} \le N. \nonumber
\end{align}
Here, constraints C1 and C2 guarantee the minimum-achievable-rate requirements, $R_{\mathrm{TH},j}^\mathrm{UL}$ and $R_{\mathrm{TH},k}^\mathrm{DL}$, for UL UT $j$ and DL UT $k$, respectively. Constraint C3 is the non-negative power constraint for UL UT $j$. Constraint C4 restrains the finite movement regions for MAs. Constraints C5 and C6 ensure the minimum inter-MA distance, $D$, for practical implementation. The total DL transmit power minimization is given by
\begin{equation}\label{P2}
	\mathop {\mathrm{minimize} }\limits_{{p_j},{\mathbf{w}_k}, \mathbf{t},\mathbf{r} } \quad \sum\limits_{k \in \mathcal{K}} \left\| \mathbf{w}_k \right\|_2^2 \quad \text{s.t.} \quad {\text{C1-C6}}.
\end{equation}

\textit{Remark 1:} The two minimization problems \eqref{P1} and \eqref{P2} are in conflict with each other. The DL signal causes significant SI, which degrades the quality of the UL reception. To satisfy the UL minimum-achievable-rate requirements, the UL UTs must increase their transmit powers, introducing substantial CCI to the DL UTs. Hence, the FD satellite has to increase its transmit power to meet the DL minimum-achievable-rate requirements, which, in turn, causes stronger SI and leads to an escalating increase in total UL and DL transmit powers.

To study the trade-off between the conflicting system design objectives, we formulate a multiobjective optimization problem using the weighted Tchebycheff method \cite{MOOP2}, i.e.,
\begin{equation}\label{P3}
	\mathop {\mathrm{minimize} }\limits_{{p_j},{\mathbf{w}_k}, \mathbf{t},\mathbf{r} } \mathop {\max }\limits_{i \in \left\{ {1,2} \right\}} \left\{ {{\lambda _i}\left( {\frac{{{T_i} - T_i^*}}{{\left| {T_i^*} \right|}}} \right)} \right\} \quad \text{s.t.} \quad {\text{C1-C6}}, 
\end{equation}
where ${T_1}=\sum\nolimits_{j \in \mathcal{J}} p_j$, ${T_2}=\sum\nolimits_{k \in \mathcal{K}} \left\| \mathbf{w}_k \right\|_2^2$, $T_i^*$ is the corresponding optimal objective value\footnote{$T_1^*$ and $T_2^*$ are treated as constants to explore the trade-off among objective functions in problems \eqref{P1} and \eqref{P2}. By setting $T_1^* = {\iota _1}$ and $T_2^* = {\iota _2}$, where $0<\iota _1,\iota _2<\infty$, we can recover solutions for problems \eqref{P1} and \eqref{P2}, respectively. However, this does not imply that the optimal values for problems \eqref{P1} and \eqref{P2} are exactly $\iota _1$ and $\iota _2$.}, and $\lambda _i \ge 0$, $\sum\nolimits_{i \in \left\{ {1,2} \right\}}\lambda _i=1$ is a weight that reflects the priority of objective $i$ compared to the other objectives. Since problem \eqref{P3} encompasses problems \eqref{P1} and \eqref{P2}, we concentrate on solving problem \eqref{P3} in the next section.
\section{Proposed Solution} \label{3}
To effectively address the non-convex problem \eqref{P3}, we propose a two-loop PSO-based algorithm in this section.
\subsection{Inner-Loop: Optimize UL Transmit Power and DL Beamformers}
In the inner-loop, the UL transmit power and DL beamformers are optimized based on the given MA positions determined by each particle. Defining $\gamma^\mathrm{UL}_{\mathrm{TH},j}\triangleq 2^{R^\mathrm{UL}_{\mathrm{TH},j}}-1$, $\gamma^\mathrm{DL}_{\mathrm{TH},k}\triangleq 2^{R^\mathrm{DL}_{\mathrm{TH},k}}-1$, and an auxiliary optimization variable as $\tau$\footnote{Note that we have $\mathbf{W}_k=\mathbf{w}_k \mathbf{w}_k^H \in {\mathbb{C}^{M \times M}}$.}, we rewrite problem \eqref{P3} as
\begin{align}\label{min_tao}
	&\mathop {\mathrm{minimize} }\limits_{{p_j},{\mathbf{W}_k},\tau}\quad \tau  \\
	&\text{s.t.} \quad {\text{C1-C3}},\quad \text{C7}: \mathbf{W}_k \succeq \mathbf{0}, \quad\forall k \in \mathcal{K},\nonumber\\
	&\hspace{2.2em} \text{C8}: \mathrm{Rank}\left\lbrace\mathbf{W}_k \right\rbrace =1,\quad \forall k \in \mathcal{K}, \nonumber\\
	&\hspace{2.2em} \text{C9}: {{\lambda _i}\left( {\frac{{{T_i} - T_i^*}}{{\left| {T_i^*} \right|}}} \right)} \le \tau,\quad \forall i \in \left\{ {1,2} \right\}.\nonumber
\end{align}
Define $\mathbf{P} \triangleq \mathrm{diag}\left(  {{p_1}, \cdots ,{p_J}} \right) $, where $\mathrm{diag}\left({{p_1}, \cdots ,{p_J}} \right) $ represents a diagonal matrix whose diagonal elements are $\left\lbrace {{p_1}, \cdots ,{p_J}} \right\rbrace $, an auxiliary variable as $\delta _k$, and the collections of the actual CCI channels, estimated CCI channels, and CCI estimation errors of DL UT $k$ as ${\mathbf{c}_k} \triangleq {\left[ {{c_{1k}}, \cdots ,{c_{Jk}}} \right]^T}$, ${{\hat {\mathbf{c}}}_k} \triangleq {\left[ {{{\hat c}_{1k}}, \cdots ,{{\hat c}_{Jk}}} \right]^T}$, and $\Delta {\mathbf{c}_k} \triangleq {\left[ {\Delta {c_{1k}}, \cdots ,\Delta {c_{Jk}}} \right]^T}$, respectively. Thus, \eqref{CCI} can be rewritten as
\begin{equation}
	{\mathbf{c}_k} = {{\hat {\mathbf{c}}}_k} + \Delta {\mathbf{c}_k}, \quad{\mathbf{\Omega}_k}  \triangleq \left\{ {{\mathbf{c}_k}:\left\| {\Delta {\mathbf{c}_k}} \right\|_2 < \varepsilon _k} \right\},
\end{equation}
where $\varepsilon _k^2 = \sum\nolimits_{j \in J} {\varepsilon _{jk}^2}$. Since constraint C2 involves infinite inequality constraints, we transform C2 into a linear matrix inequality constraint using the following proposition.

\textit{Proposition 1:} Constraint C2 can be equivalently expressed as
\begin{equation}
	\overline{\text{C2}}:\left[ {\begin{array}{*{20}{c}}
			{{\delta _k}{\mathbf{I}_J} - \mathbf{P}}&{ - \mathbf{P}{{\hat{\mathbf{c}}}_k}}\\
			{ - \hat{\mathbf{c}}_k^H\mathbf{P}}&{{\chi _k}}
	\end{array}} \right] \succeq \mathbf{0},
\end{equation}
where $\mathbf{I}_J$ denotes an identity matrix of size $J \times J$, and
\begin{align}
	{\chi _k} =  &- {\delta _k}\varepsilon _k^2 - \hat{\mathbf{c}}_k^H\mathbf{P}{{\hat{\mathbf{c}}}_k} \nonumber\\
	&- \sum\limits_{i \in \mathcal{K}\backslash \left\{ k \right\}} {\mathrm{Tr}\left\{ {{\mathbf{W}_i}{\mathbf{H}_k}} \right\}}+ \frac{{\mathrm{Tr}\left\{ {{\mathbf{W}_k}{\mathbf{H}_k}} \right\}}}{{\gamma _{\mathrm{TH},k}^\mathrm{DL}}} - \sigma _{\mathrm{DL},k}^2.
\end{align}
\begin{IEEEproof}
	Please refer to \cite[Lemma 1]{MOOP2}.
\end{IEEEproof}

Besides, to address the non-convex rank-one constraint C8, semidefinite relaxation (SDR) is applied to remove it. Then, problem \eqref{min_tao} becomes a convex semidefinite programming (SDP) and is given by
\begin{align}\label{max_tau}
	&\mathop {\mathrm{minimize} }\limits_{{p_j},{\mathbf{W}_k},\tau,\delta _k}\quad \tau  \\
	&\text{s.t.} \quad \text{C1},\overline{\text{C2}},\text{C3},\text{C7},\text{C9},\quad\text{C10}: \delta _k \ge 0,\quad \forall k \in \mathcal{K}. \nonumber
\end{align}
Problem \eqref{max_tau} can be solved optimally with standard convex optimization tools. Furthermore, the tightness of the rank relaxation is verified in \cite[Theorem 1]{MOOP1}.
\subsection{Outer-Loop: Optimize MA Positions}
The optimized objective value of problem \eqref{max_tau} is incorporated into the fitness function of the outer-loop PSO algorithm for optimizing the MA positions. The PSO algorithm begins by initializing the positions and velocities of $Z$ particles as ${\mathbf{U}^{\left( 0 \right)}} = \left\{ {\mathbf{u}_1^{\left( 0 \right)}, \cdots ,\mathbf{u}_Z^{\left( 0 \right)}} \right\} \in {\mathbb{R}^{2\left( M+N\right)  \times Z}}$ and ${\mathbf{V}^{\left( 0 \right)}} = \left\{ {\mathbf{v}_1^{\left( 0 \right)}, \cdots ,\mathbf{v}_Z^{\left( 0 \right)}} \right\} \in {\mathbb{R}^{2\left( M+N\right)  \times Z}}$, respectively. Each particle represents a potential solution, where its position vector is $\mathbf{u}_z^{\left( 0 \right)} = {\left[ { {\mathbf{t}}^{\left(0 \right) T}_z},{{\mathbf{r}}^{\left(0 \right) T}_z} \right]^T} \in {\mathbb{R}^{2\left( M+N\right)  \times 1}}$ ($1 \le z \le Z$). The initial positions and velocities are randomly generated within the feasible regions defined by constraints C4-C6. For each particle, the initial position is considered its personal best position, $\mathbf{u}_{\mathrm{pbest},z}$. The global best position, $\mathbf{u}_{\mathrm{gbest}}$, is determined based on the particle with the minimum fitness value. Once the initialization is complete, the details of the PSO algorithm are provided below.
\subsubsection{Update Positions and Velocities}
Denote the maximum number of iterations as $Q$. The velocity of each particle in the $q$-th iteration ($1 \le q \le Q$) is updated as
\begin{align} \label{v}
	\mathbf{v}_z^{\left( q \right)}  = \omega  \mathbf{v}_z^{\left( q-1 \right)}& + {\alpha_1}{\mathbf{e}_1} \odot \left( {{\mathbf{u}_{\mathrm{pbest},z}} - \mathbf{u}_z^{\left( {q - 1} \right)}} \right) \nonumber\\
	& + {\alpha_2}{\mathbf{e}_2} \odot \left( {{\mathbf{u}_\mathrm{gbest}} - \mathbf{u}_z^{\left( {q - 1} \right)}} \right) ,
\end{align}	
where $\odot$ represents the Hadamard product. $\omega$ is the inertia weight, which decreases linearly over the iterations within $\left[\omega_\mathrm{min},\omega_\mathrm{max} \right] $, i.e., $\omega = \omega_\mathrm{max}- \left(  \omega_\mathrm{max} - \omega_\mathrm{min}\right)\frac{ q}{Q}$. $\alpha_1$ and $\alpha_2$ are the learning factors. $\mathbf{e}_1$ and $\mathbf{e}_2$ are the random vectors used to avoid undesired local optimal solutions, and their elements are uniformly distributed within $\left[0,1\right]$. Then, the position of each particle in the $q$-th iteration is updated as
\begin{equation} \label{u}
	\mathbf{u}_z^{\left( q \right)} = \mathcal{B}\left\{ {\mathbf{u}_z^{\left( {q - 1} \right)} + \mathbf{v}_z^{\left( q \right)}} \right\},
\end{equation}
where $\mathcal{B}\left\lbrace \cdot \right\rbrace $ is a projection function that guarantees constraint C4 \cite{Ding2}.
\subsubsection{Update Personal and Global Best Positions}
To evaluate the position of each particle, we define a fitness function as
\begin{equation} \label{fitness}
	\mathcal{F}\left( {\mathbf{u}_z^{\left( q \right)}} \right) = \tau\left( {\mathbf{u}_z^{\left( q \right)}} \right) + \beta \xi\left( {\mathbf{u}_z^{\left( q \right)}} \right),
\end{equation}
where $\tau\left( {\mathbf{u}_z^{\left( q \right)}} \right)$ is the objective value obtained after solving problem \eqref{max_tau} with the MA positions given by $\mathbf{u}_z^{\left( q \right)}$, $\beta$ is a positive penalty factor, and $\xi\left( {\mathbf{u}_z^{\left( q \right)}} \right)$ is a counting function that returns the number of MAs violating constraints C5 and C6 at position ${\mathbf{u}_z^{\left( q \right)}}$. Given that the minimum fitness value is assumed to correspond to the best position, the penalty term can push the global best position to satisfy constraints C5 and C6. Then, the personal and global best positions are updated if the fitness value at the current position is lower than their respective minimum fitness values. After $Q$ iterations, the optimized MA positions, $\mathbf{u}_\mathrm{gbest}$, UL transmit powers, $p_j$, and DL beamformers, $\mathbf{w}_k$, are obtained. 
\subsection{Convergence and Complexity Analysis}
Since only the position with a lower fitness value is selected as the global best position, the global best fitness value is non-increasing during the iterations, i.e., $\mathcal{F}( \mathbf{u}_\mathrm{gbest}^{\left( q\right) }) \le \mathcal{F}( \mathbf{u}_\mathrm{gbest}^{\left( q-1\right) })$.
Besides, the transmit power is lower-bounded by zero, i.e., we have $\mathcal{F}( \mathbf{u}_\mathrm{gbest}^{\left( q\right) }) > 0$. Therefore, the convergence of the proposed algorithm is guaranteed. Moreover, the convergence is verified by the simulations in Section \ref{4}.

The computational complexity of the proposed algorithm primarily stems from the iterations of the PSO algorithm and the process of solving problem \eqref{max_tau} for each particle, for which the complexities are $\mathcal{O}\left( ZQ\right) $ and $\mathcal{O}\left( KM^{3.5}+J^{3.5}\right) $ \cite{MOOP2}, respectively. Hence, the overall complexity is $\mathcal{O}\left(ZQ\left(  KM^{3.5}+J^{3.5}\right)\right)  $.
\section{Simulation Results} \label{4}
In this section, the performance of the proposed scheme is evaluated through simulation results. In the simulation, we consider a LEO FD satellite with an orbit altitude of 600 kilometers (km). The UL and DL UTs are randomly distributed within the coverage areas of the beams, with the distances between them ranging from 1 km to 10 km. The carrier frequency is set to 8 GHz and the path loss exponent is set to 2.8. For the SI and UL/DL channels, we respectively adopt the channel models in \cite{Ding1} and \cite{MA3}, where the SI loss coefficient is set to $\rho=-100$ dB, and the numbers of transmit and receive SI paths are the same, i.e., $L^\mathrm{t}=L^\mathrm{r}=10$. For the CCI channel, we define the maximum normalized estimation error as $\frac{{\varepsilon _{jk}^2}}{{{{\left| {{c_{jk}}} \right|}^2}}} =5\%$. Unless otherwise stated, for the proposed communication system, we set the numbers of transmit and receive MAs to $M=N\triangleq\widetilde N=16$, the moving region sizes to $A^\mathrm{t}=A^\mathrm{r}\triangleq A=5\lambda$, the numbers of UL and DL UTs to $J=6$ and $K=2$, the minimum-achievable-rate requirements to $R^\mathrm{UL}_{\mathrm{TH},j}=0.5$ bps/Hz and $R^\mathrm{DL}_{\mathrm{TH},k}=1$ bps/Hz, the average noise powers of the FD satellite and the DL UTs to $\sigma^2_\mathrm{UL}=-110$ dBm and $\sigma^2_{\mathrm{DL},k}=-100$ dBm, the FD satellite antenna gain to 20 dBi, and the minimum inter-MA distance to $D=\frac{\lambda}{2}$. For the PSO-based algorithm, we set the parameters according to \cite{MA2,Ding2}, where the penalty factor is set to $\beta=1$.
\begin{figure}[!t]
	\centering
	\subfloat[]{\label{fit}\includegraphics[width=0.5\columnwidth]{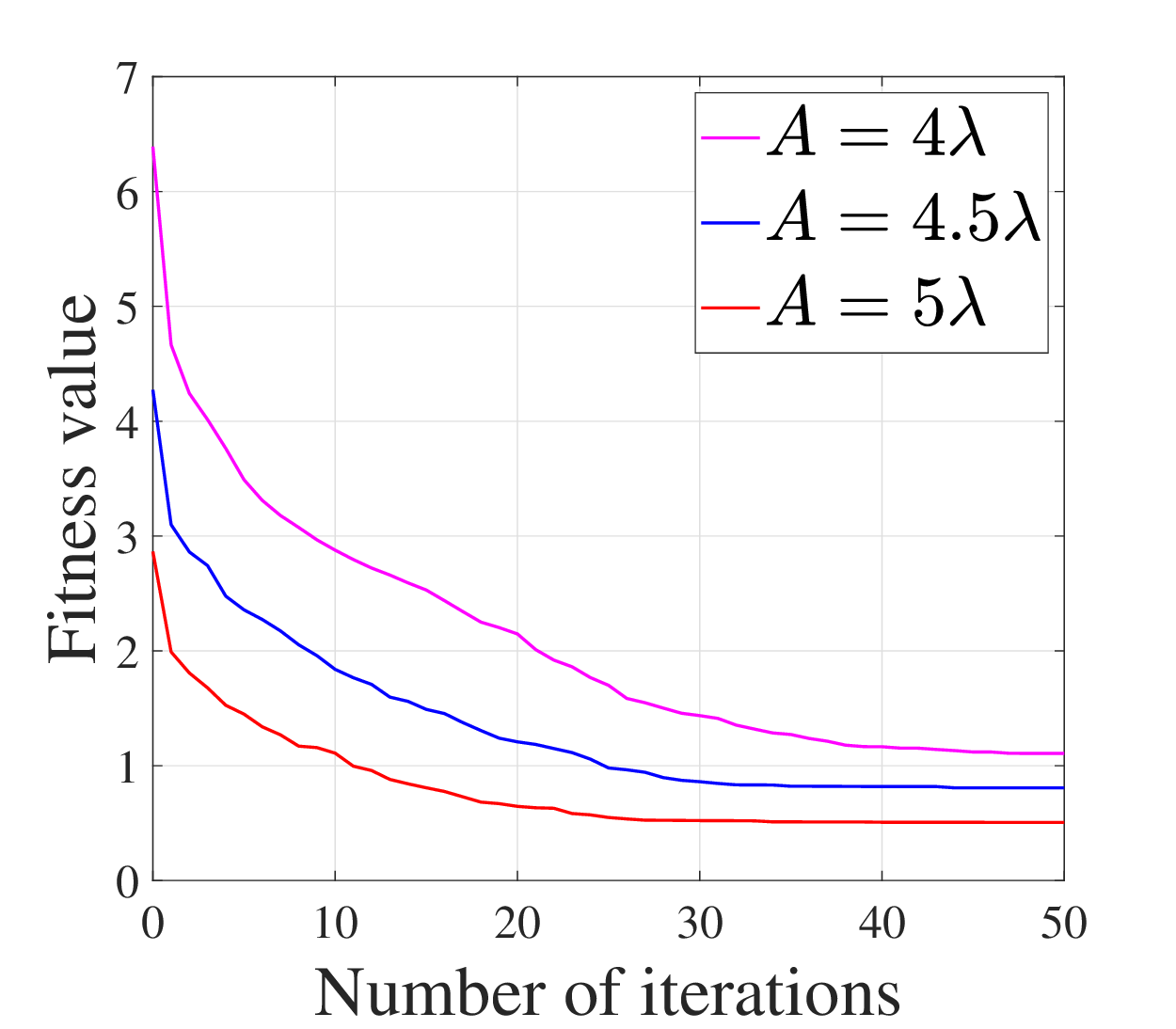}}
	\subfloat[]{\label{pen}\includegraphics[width=0.5\columnwidth]{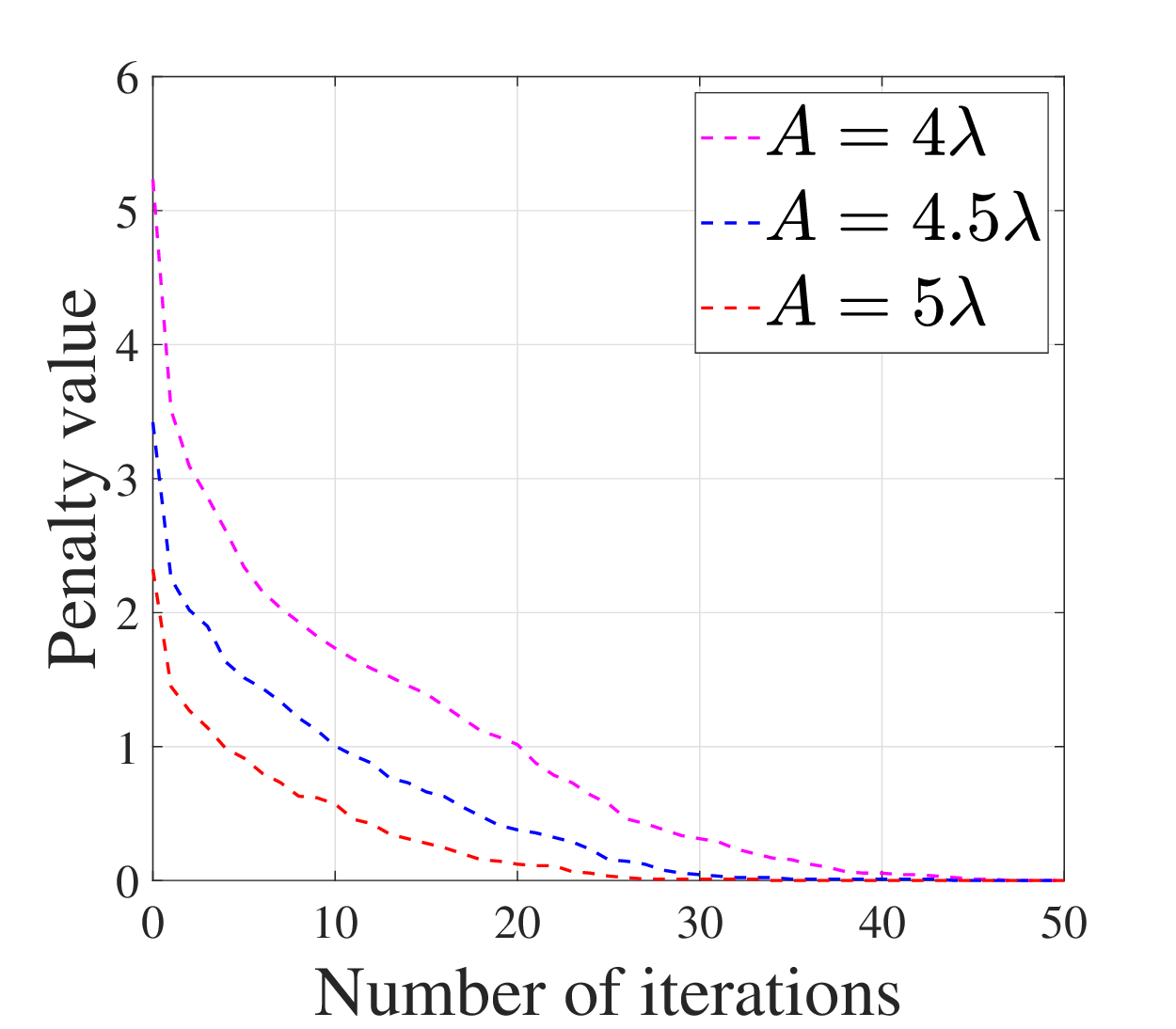}}
	\caption{Convergence evaluation of the proposed algorithm.}
	\label{conv}
\end{figure}
\begin{figure}
	\centering
	\includegraphics[width=0.6\linewidth]{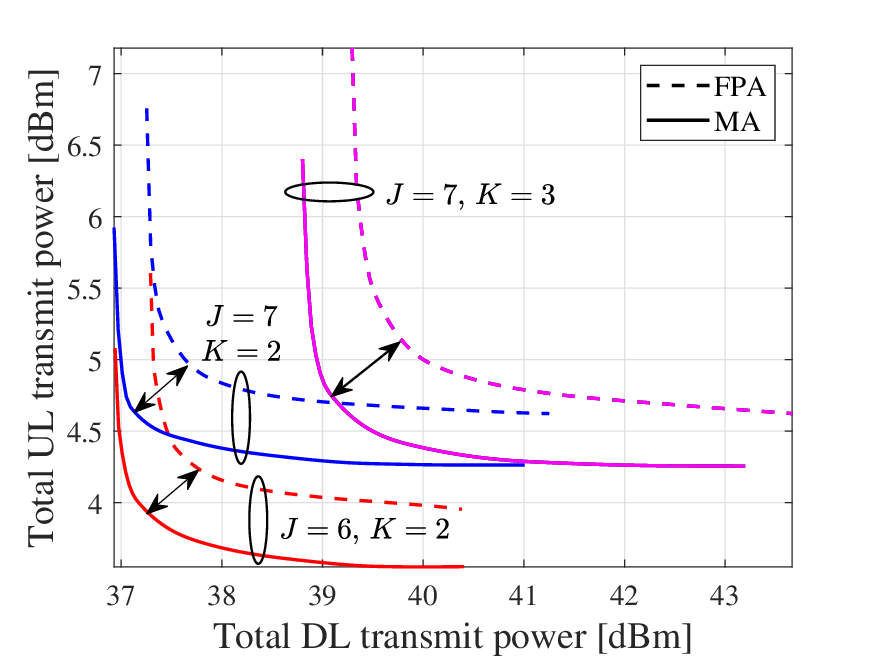}
	\caption{Average trade-off region of total UL and DL transmit powers.}
	\label{trade_off}
\end{figure}
\begin{figure}[!t]
	\centering
	\subfloat[]{\label{SI_UL}\includegraphics[width=0.5\columnwidth]{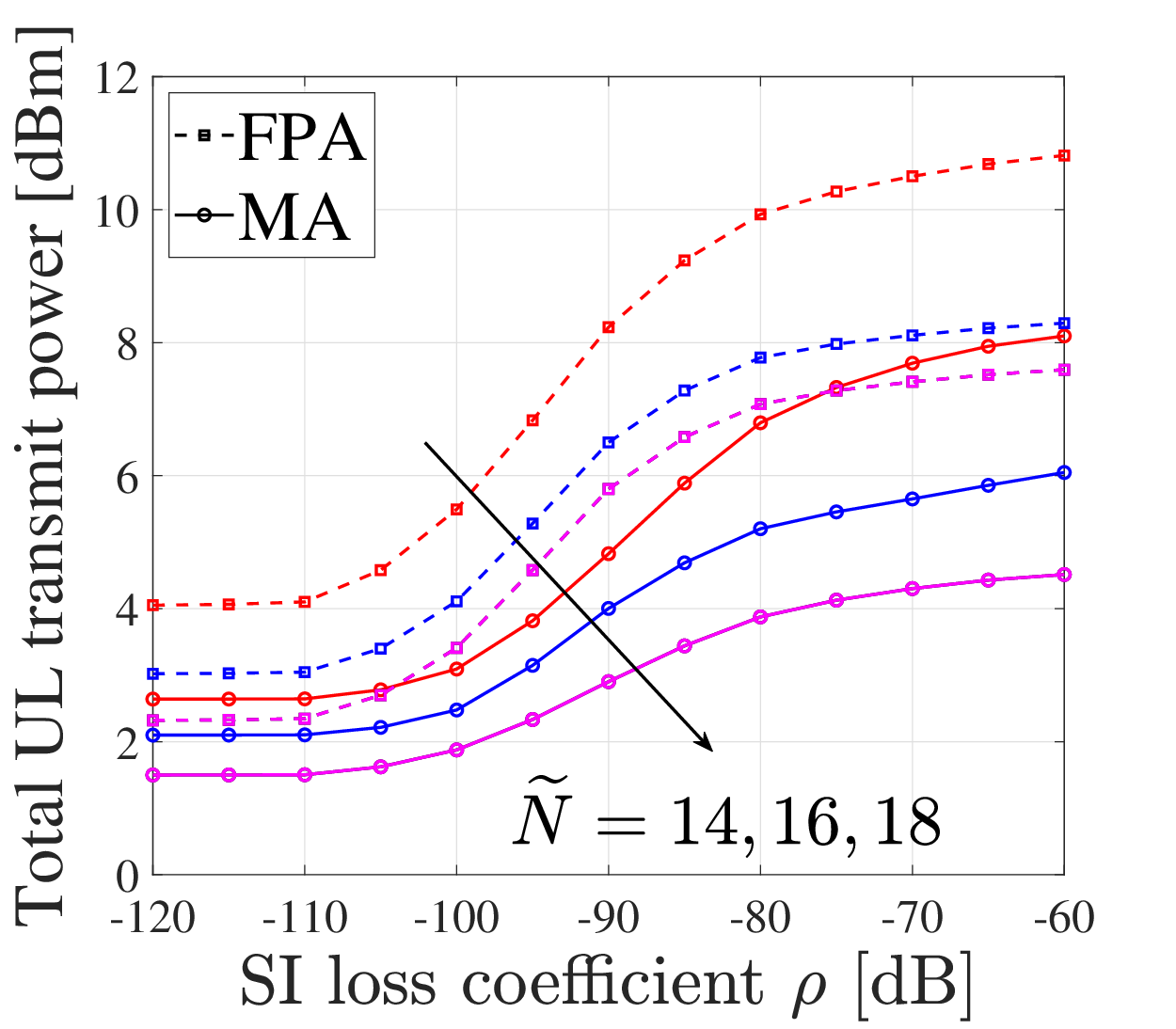}}
	\subfloat[]{\label{SI_DL}\includegraphics[width=0.5\columnwidth]{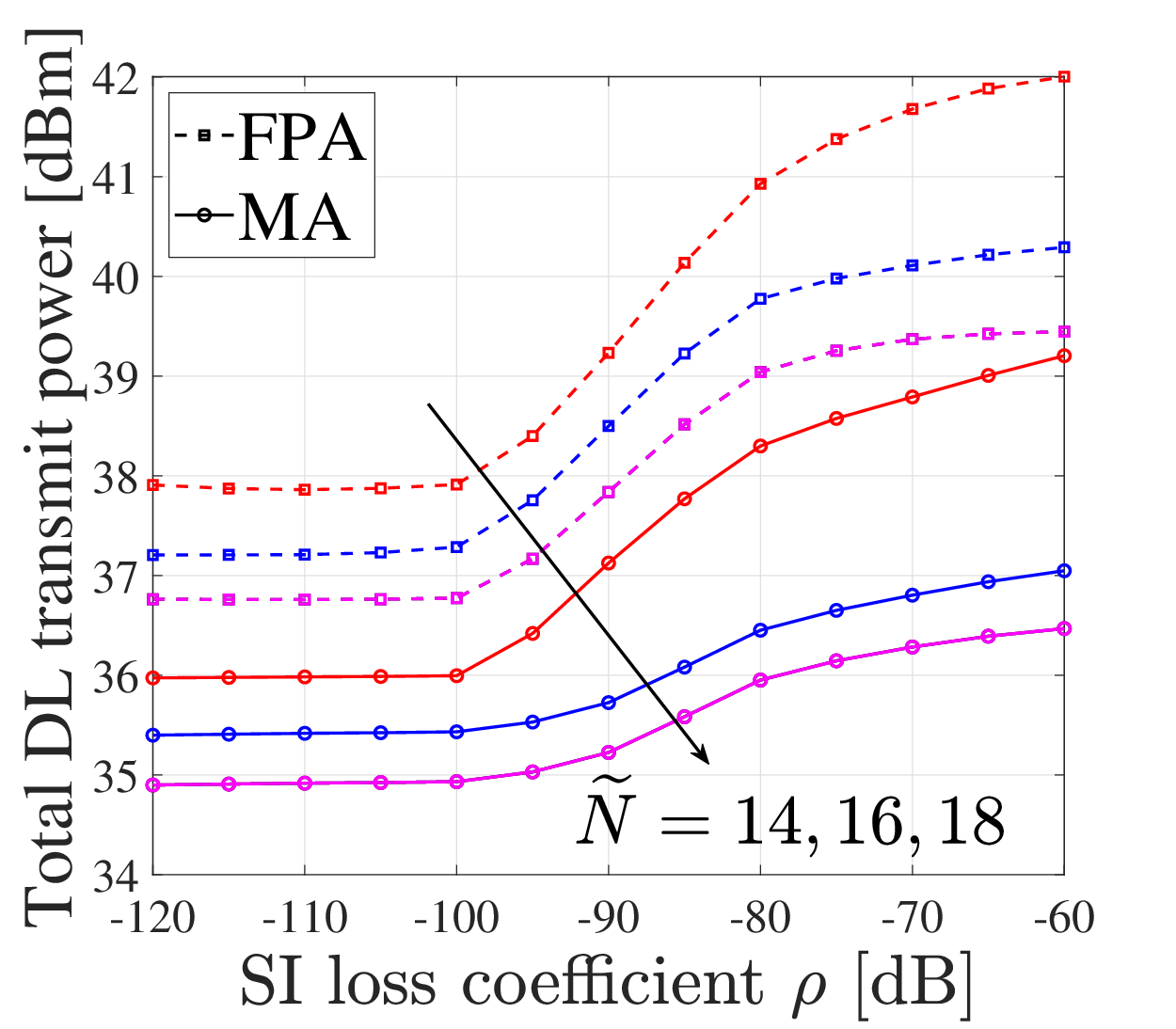}}
	\caption{Total (a) UL and (b) DL transmit powers versus SI loss coefficient.}
	\label{SI}
\end{figure}

Fig.\;\ref{conv} depicts the convergence performance of the proposed algorithm. We can see that, for different moving region sizes, both the fitness and penalty values decrease steadily with the number of iterations, reaching stable values within 50 iterations. In Fig.\;\ref{conv}\subref{fit}, the fitness value converges more quickly and achieves a lower level for larger moving region sizes, while smaller moving region sizes require more iterations to stabilize. This is because expanding the moving region sizes allows the MAs to explore wider ranges, thus facilitating a faster identification of superior MA positions. In Fig.\;\ref{conv}\subref{pen}, the penalty values ultimately converge to zero, ensuring that constraints C5 and C6 are consistently satisfied.

Fig.\;\ref{trade_off} shows the trade-off between the total UL and DL transmit powers. For comparison, we consider a baseline scheme, namely FPA, where the FD satellite is equipped with an FPA-based uniform planar array with $\widetilde N$ transmit and $\widetilde N$ receive antennas, spaced by $\frac{\lambda}{2}$. The corresponding UL and DL transmit powers are obtained by solving problem \eqref{max_tau}. The trade-off region is established by uniformly varying $\lambda _i$ with a step size of 0.01, subject to $\lambda _i \ge 0$ and $\sum\nolimits_{i \in \left\{ {1,2} \right\}}\lambda _i=1$. As shown in Fig.\;\ref{trade_off}, the flexible movements of MAs effectively alleviate the trade-off between UL and DL transmit powers and achieve comparable communication performance to the FPA scheme while utilizing lower transmit power.

Fig.\;\ref{SI} compares the total UL and DL transmit powers for systems using MAs and FPAs, with respect to the SI loss coefficient. It can be seen that, for both the MA and FPA schemes, the total UL and DL transmit powers increase with the SI loss coefficient across varying antenna numbers. This is because, with weak SI cancellation, the UL UTs have to transmit signals with higher power to maintain the communication quality, which leads to increased CCI for DL UTs. Consequently, the FD satellite must allocate more transmit power to meet the DL minimum-achievable-rate requirements. Moreover, at the same UL or DL transmit power level, the MA scheme requires lower SI cancellation capability than the FPA scheme due to antenna position optimization.

\section{conclusion}\label{5}
This letter investigated a power-efficient MA-aided FD satellite communication system, where the FD satellite is equipped with multiple transmit and receive MAs. To study the trade-off between the total UL and DL transmit powers, a PSO-based multiobjective optimization algorithm was proposed to jointly optimize the MA positions and the power allocation for UL and DL. The results showed that the MA can save the UL and DL transmit powers and reduce the required SI cancellation capability compared to conventional FPA systems, which enhances the performance and efficiency of FD satellite communication systems.\\

\newpage

\vfill

\end{document}